\documentclass[preprint,showpacs,preprintnumbers,amsmath,amssymb]{revtex4}

\usepackage{graphicx}
\usepackage{dcolumn}
\usepackage{bm}
\usepackage{epsfig}
\begin{document}

\title{Study of the effects of Pauli blocking and Pauli non-locality on the 
optical potential.}
 
\author{M. A. G. Alvarez and N. Alamanos}
\affiliation{DSM/DAPNIA/SPhN CEA-Saclay, 91191 Gif-sur-Yvette, France}
\date{\today}

\author{L. C. Chamon and M. S. Hussein}
\affiliation{Instituto de F\'{\i}sica da Universidade de S\~{a}o Paulo, 
Caixa Postal 66318, 05315-970 S\~{a}o Paulo, SP, Brazil}
\date{\today}
\begin{abstract}
Elastic scattering angular distributions for systems with reduced mass between 3
and 34 and energies varying between 25 and 120 MeV/nucleon were analyzed. The 
stable $^4$He, its exotic partner $^6$He, and the weakly bound $^{6,7}$Li 
nuclei were included as projectiles in the systematics. Optical model data 
analyzes were performed with an adjustable factor of normalization included in 
the imaginary part of the potential. These analyzes indicated a reduction 
of absorption for systems with small reduced masses that was detected due to 
the refractive nature of the scattering by light systems.
\end{abstract}


\maketitle

\section{\label{sec:level1}Introduction}

The elastic scattering of light heavy-ions at intermediate energies has clearly
demonstrated the sensitivity at large angles to the underlying optical potential
through the "appearance" in the angular distribution of Airy oscillations
associated with nuclear rainbow scattering \cite{nova4}. Such studies pin down 
several aspects of the optical potential, usually constructed through the 
double-folding procedure. Among these aspects, we mention the degree of 
non-locality, genuine energy-dependence, the density dependence of the 
effective nucleon-nucleon G-matrix, etc.

Recently, the elastic scattering of halo-type nuclei, such as $^{11}$Li,
$^6$He, $^{11}$Be, $^{19}$C, at intermediate energies has been studied. In such
cases, due to the low beam intensity, it is rather difficult to cover the Airy 
region. Thus, at most, one is bound to extract from the small angle, near/far 
interference, region, information about the strength of the coupling to the
break-up channel. The energy dependence associate with non locality of the local
equivalent optical potential has a paramount importance in such studies.

In this work, we discuss the elastic scattering of stable, weakly bound and
exotic nuclei on a variety of targets, both light and heavy, in order to assess 
the energy-dependence. For this purpose, we use the S\~{a}o Paulo potential and 
the Lax interaction discussed in details in Refs. \cite{4,10,15}. In section II, 
we give an account of the optical potential and its energy-dependence. In 
section III, we present the data analysis. Finally, in section IV, we present a 
summary and concluding remarks.

\section{\label{sec:level2}The optical potential}

Two different phenomena, called Pauli non-locality (PNL) and Pauli blocking 
(PB), are important to understand the energy-dependence of the optical potential
for heavy-ion systems. The PNL arises from quantum exchange effects and has been 
studied in the context of neutron-nucleus \cite{1}, alpha-nucleus \cite{2} and 
heavy-ion \cite{4,10,3,5,6,7,8,9,11,12,13,14} collisions. The nonlocal 
interaction has been used in the description of the elastic scattering process 
through an integro-differential equation \cite{4,10,1}. It is possible to 
define a local-equivalent potential which, within the usual framework of the 
Schroedinger differential equation, reproduces the results of the
integro-differential approach \cite{10,1}. In the case of heavy-ion systems, 
the real part of the local-equivalent interaction is associated to the 
double-folding potential ($V_F$) through \cite{10}:
\begin{equation}
V_{N}(R) = V_{F}(R)e^{-4v^2/c^2}
\end{equation}
where {\em $c$\/} is the speed of light and {\em $v$\/} is the local relative 
velocity between the two nuclei. This model is known as S\~{a}o Paulo potential. 
The velocity/energy-dependence of the potential is very important to account for
the data from near-barrier to intermediate energies 
\cite{4,10,3,5,6,7,8,9,11,12,13}. Eq. (1) describes the effect of the PNL on the 
real part of the potential, but the local-equivalent potential that arises from 
the solution of the corresponding integro-differential equation indicates that 
the exchange correlation also affects the imaginary part of the optical 
potential \cite{4}.

Another model used in the analyzes of elastic scattering data is the Lax 
interaction \cite{15}, which is the optical limit of the Glauber high-energy 
approximation \cite{16}. The Lax interaction is essentially a zero range 
double-folding potential used for both the real and imaginary parts of the 
optical potential. Similar to the S\~{a}o Paulo potential, the Lax interaction 
is also 
dependent on the nuclear densities and may be expressed in terms of the relative 
velocity between the two nuclei. The imaginary part of the Lax interaction is
thus written as:
\begin{equation}
W(R) = -\frac{1}{2} \hbar v \int \sigma_{T}^{NN}(v) \rho_{1}(\vec{r}) 
\rho_{2}(\vec{r}-\vec{R}) \; d\vec{r}
\end{equation}
where $\sigma_{T}^{NN}(v)$ is an energy-dependent spin-isospin-averaged total 
nucleon-nucleon cross-section. Eq. (2) has been derived from multiple-scattering 
theories and should be valid for stable (non-exotic) nuclei at high energies. 
For lower energies, Eq. (2) must be corrected in order to take into account the 
closure of the phase space due to the Pauli exclusion principle. This 
phenomenon, known as Pauli blocking (PB), can be simulated in Eq. (2) by 
introducing a further dependence of $\sigma_{T}^{NN}$ on the densities of the 
nuclei \cite{15}. Usually, the PB is expected to be essential at small 
internuclear distances due to the corresponding large overlap of the nuclei. 
Since the PB can distort significantly the densities in the overlap region, it 
should affect both the real and imaginary parts of the optical potential. 

In this work, we have assumed these two semi-phenomenological models 
for the real (Eq. 1) and imaginary (Eq. 2) parts of the optical potential. We 
have analyzed several elastic scattering angular distributions for systems with 
reduced mass between 3 and 34. As extensively discussed in Ref. \cite{10}, Eq.
(1) can be used in several different frameworks that provide very similar 
results in data analyzes. In the present work, we use Eq. (1) within the 
zero-range approach for the effective nucleon-nucleon interaction
($v_{nn}(\vec{r})=V_0 \delta(\vec{r})$) with the matter densities assumed in 
the folding calculations (see \cite{10}). This approach is equivalent \cite{10} to 
the more usual procedure of using the M3Y nucleon-nucleon interaction with the 
nucleon densities of the nuclei. For $\sigma_{T}^{NN}$ in Eq. (2), we have 
interpolated values from the corresponding experimental results of Ref. 
\cite{nova2}.

\section{\label{sec:level3}Data Analysis}

Table 1 presents all systems that have been analyzed in the present work. The
data have been obtained from Refs. \cite{20,21,22,23,24,25,26,27,28}. 
Eqs. (1) and (2) involve the folding of the nuclear densities. In an earlier 
paper \cite{10}, we presented an extensive systematics of heavy-ion densities. 
In that work, the Fermi distribution was assumed to describe the densities. The 
systematics indicates that the radii of the matter distributions are well 
represented by:
\begin{equation}
R_0=1.31 A^{1/3} - 0.84 \; \mbox{fm}.
\end{equation}
where $A$ is the number of nucleons of the nucleus. The densities present an
average diffuseness value of $a=0.56$ fm. Owing to specific nuclear structure
effects (single particle and/or collective), the parameters $R_0$ and $a$ show
small variations around the corresponding average values throughout the periodic
table. In the present work, we have assumed Eq. (3) for all nuclei and allowed
$a$ to vary around its average value in order to obtain the best data fits. The
only exception is the $^4$He nucleus for which the shape of the corresponding
matter density was assumed to be similar to the charge density obtained from
electron scattering experiments \cite{19}. Of course, the use of a Fermi-type
density is not well justified for light, both stable and unstable, nuclei.
However, we decided to use this universal form in order to assess the adequacy
and limitations of our model. The values obtained for the diffuseness of the 
nuclei are shown in Table 2. In a consistent manner, these values present very 
small variations around the average value obtained in the previous systematics: 
$a=0.56$ fm. We emphasize the much greater value obtained for the diffuseness 
of the exotic $^6$He in comparison with its partner $^4$He. Indeed, the 
diffuseness of the $^6$He is comparable with the values obtained for heavy-ions. 
Similar results have already been observed in other works \cite{11,12}. Table 2 
also presents the root-mean-square (RMS) radii for the matter densities and for 
charge distributions extracted from electron scattering experiments \cite{19}. 
The RMS radii for the matter densities agree with the values for charge 
distributions within about 5\%, except for $^{12}$C where a difference of 10\% 
was found.

For the imaginary part of the optical potential we have adopted Eq. (2), without
PB, multiplied by a factor of normalization $N_I$. The corresponding predictions 
for some elastic scattering angular distributions are shown in Figs. (1-3). In 
these figures, the dashed lines represent the predictions with $N_I=1$ while the 
solid ones correspond to the results obtained considering $N_I$ as a free 
parameter. 

The best fit values obtained for $N_I$ are presented in Table 1 and Fig. 4. A 
strong reduction of absorption is observed for systems with small reduced mass. 
As already commented in section I, PNL and PB affect both real and imaginary 
parts of the optical potential. The detected reduction of absorption could 
partially arises from the effect of the PB. In order to illustrate this point, 
Fig. (5 - Bottom) shows $W(R)$ obtained from Eq. (2), for the 
$^{12}$C + $^{12}$C system in two different energies, with (solid lines) and 
without (dashed lines) including the effect of PB. The calculation of $W(R)$ 
with PB was performed considering a density-dependent nucleon-nucleon total 
cross section according Refs. \cite{15,20}. Clearly, the blocking reduces 
$W(R)$ in an internal region of distances and almost no effect is observed in 
the surface region. This behavior is connected with the large overlap of the 
densities for small distances that makes the blocking very effective. In Fig. 
(5 - Top) we show the results of a notch test, where we have included a spline 
with Gaussian shape in the imaginary potential and calculated the variation of 
the chi-square as a function of the position of this perturbation. This test 
has the purpose of determining the region of sensitivity that is relevant for 
the elastic scattering process. Just as a guide, the position of the s-wave 
barrier radius is also indicated in the same figure. In the region of 
sensitivity, the imaginary potential with PB is in average less intense than 
the result without blocking. This fact probably is connected with the value 
$N_I \approx 0.6$ obtained for the $^{12}$C + $^{12}$C system. Still in Fig. (5) 
one can observe that the region of sensitivity is more internal for the higher 
energy. However, the difference between $W(R)$ with and without blocking is 
smaller for the higher energy. Probably, this two effects cancelate each other 
and one obtains approximately the same $N_I$ value for the two energies 
(see Table 1). In Fig. (6) we present the reaction cross sections for the 
$^{12}$C + $^{12}$C system in a very wide energy range (from Refs. 
\cite{24,29,30,31,32,33,34}). The lines represent the predictions obtained with 
$N_I=1$ and $N_I=0.6$, where the smaller value was obtained from the elastic 
scattering data fits. Clearly the value $N_I=0.6$ also provides a better 
reproduction of the reaction data in the energy region studied in this work: 
$25 \le E \le 120$ MeV/nucleon that corresponds to $300 \le E_{Lab} \le 1440$ 
MeV for $^{12}$C + $^{12}$C.

In Fig. (7), we present the notch test for systems with different reduced
masses, but for approximately the same bombarding energy. Again as a guide, the 
positions of the corresponding s-wave barrier radii are indicated in the figure. 
For the heaviest system, the region of sensitivity is close to the barrier 
radius and therefore it is in the surface region. The lighter systems present 
sensitivity regions much more internal in comparison with the corresponding 
barrier radii. In fact, it is well known that the scattering between light
heavy-ions probes more efficiently the internal internuclear distance region 
\cite{nova4} and the present results of the notch test just confirms this 
point. Thus, the simple approach of using Eqs. (1) and (2), with $N_I=1$, fails
for lighter systems that are sensitive to inner distances. The discussion about 
Fig. 5 clearly shows that the effect of PB on the imaginary part of the
potential should be partially responsible by this behavior of light systems, 
but the present analysis can not discern if such behavior is also connected with
effects of PNL on the imaginary part and/or even of PB on the real part of the
optical potential. On the other hand, considering our results for heavier
systems, the analysis clearly indicates that the surface region of the optical 
potential is well represented by Eq. (1), real part that includes the PNL 
effect, and Eq. (2), imaginary part without the PB effect.  

In order to check the consistency between the present and earlier works, we have
calculated the volume integral of the real part of the potential, Eq. 4, and the
reaction cross sections that are connected with the absorptive part of the
potential. 
\begin{equation}
J_R= \frac{4\pi}{A_1 A_2} \int V(R) R^2 dR
\end{equation}
The values obtained for $J_R$ and $\sigma_R$ (see Table 1) are similar to those 
of earlier works (from Refs. 
\cite{20,21,22,23,24,26,28,37,38,39,40,42,43,44}). In the present 
systematics we have included the weakly bound $^{6,7}$Li nuclei and also the 
exotic $^{6}$He. In some works, due to the 
break-up process or effects of the halo, these projectiles have been pointed out 
as responsible by a different behavior in comparison with nuclear reactions 
involving only normal stable nuclei. In fact, our systematics for $N_I$ also 
indicates a slightly greater absorption for systems involving $^{6,7}$Li in 
comparison with other systems with similar reduced masses (see Fig. 4). On the 
other hand, as already commented, the diffuseness obtained in this work for 
$^6$He is much greater than the value extracted for $^4$He from electron 
scattering experiments.

\section{\label{sec:level4}Summary and Conclusions}

In summary, we have analyzed elastic scattering angular distributions for 
several systems. The real part of the optical potential was assumed to be 
energy-dependent due to the PNL that arises from quantum exchange effects. For 
the imaginary part, we have adopted the Lax interaction that also presents an 
energy-dependence very well established and connected with the total 
nucleon-nucleon cross section. In the imaginary part, we have also considered a 
factor of normalization with the aim of simulating the effect of PB, which 
arises from the exclusion principle and prevent scattered nucleons to occupy 
filled states. The notch test indicated that lighter systems present greater 
sensitivity to the internal region of internuclear distances, where PB is 
expected to be essential due to the large overlap of the nuclei. For heavier 
systems it was possible to obtain a good data description with $N_I=1$. This 
result indicates that, in the surface region, the PNL and PB does not 
significantly affect the imaginary part of the potential, while Eq. (1) 
correctly describes the PNL effect on the real part. For lighter systems, 
however, reasonable accounts of the data were obtained only considering $N_I$ as 
a free parameter. This result is compatible with the expected reduction of the 
absorption due to PB and also with the fact that light nuclei have a smaller
density of states. The present analysis, however, can not discern if such 
behavior is also connected with effects of PNL on the imaginary part and/or 
even of PB on the real part of the optical potential.

\begin{acknowledgments}
This work was partially supported by Financiadora de Estudos e Projetos (FINEP), 
Funda\c{c}\~{a}o de Amparo \`a Pesquisa do Estado de S\~ao Paulo (FAPESP), and 
Conselho Nacional de Desenvolvimento Cient\'{\i}fico e Tecnol\'ogico (CNPq).
\end{acknowledgments}

\newpage

\begin{figure}
\vspace{0.0cm}
\hspace{-10.0cm}
\includegraphics{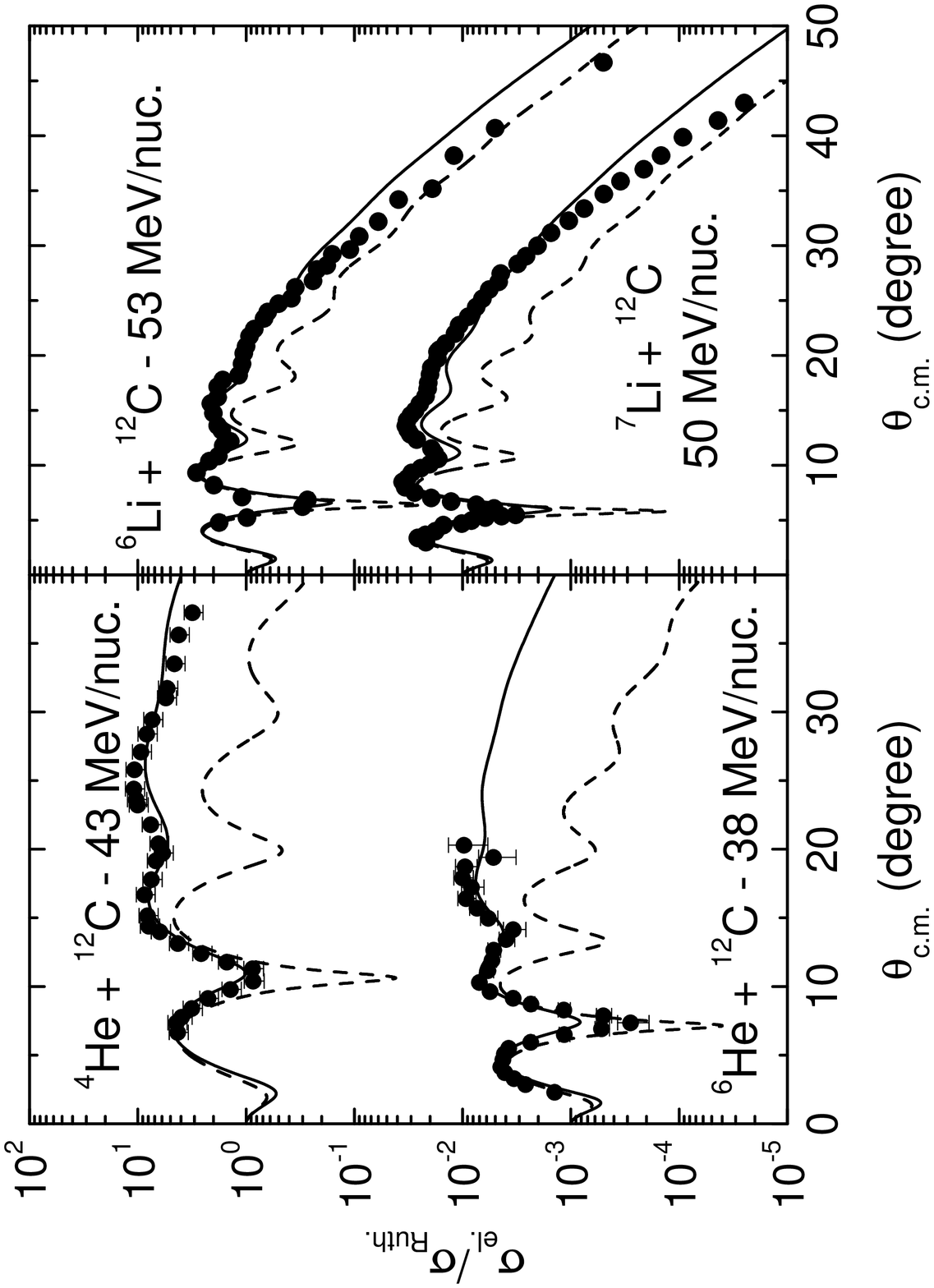}
\vspace{7.0cm}
\caption{Elastic scattering angular distributions for the 
$^{4,6}$He,$^{6,7}$Li + $^{12}$C systems. The lines correspond to optical model 
predictions with (solid lines) or without (dashed lines) including a factor of 
normalization in the imaginary part of the optical potential.}
\end{figure}

\begin{figure}
\vspace{0.0cm}
\hspace{-10.0cm}
\includegraphics{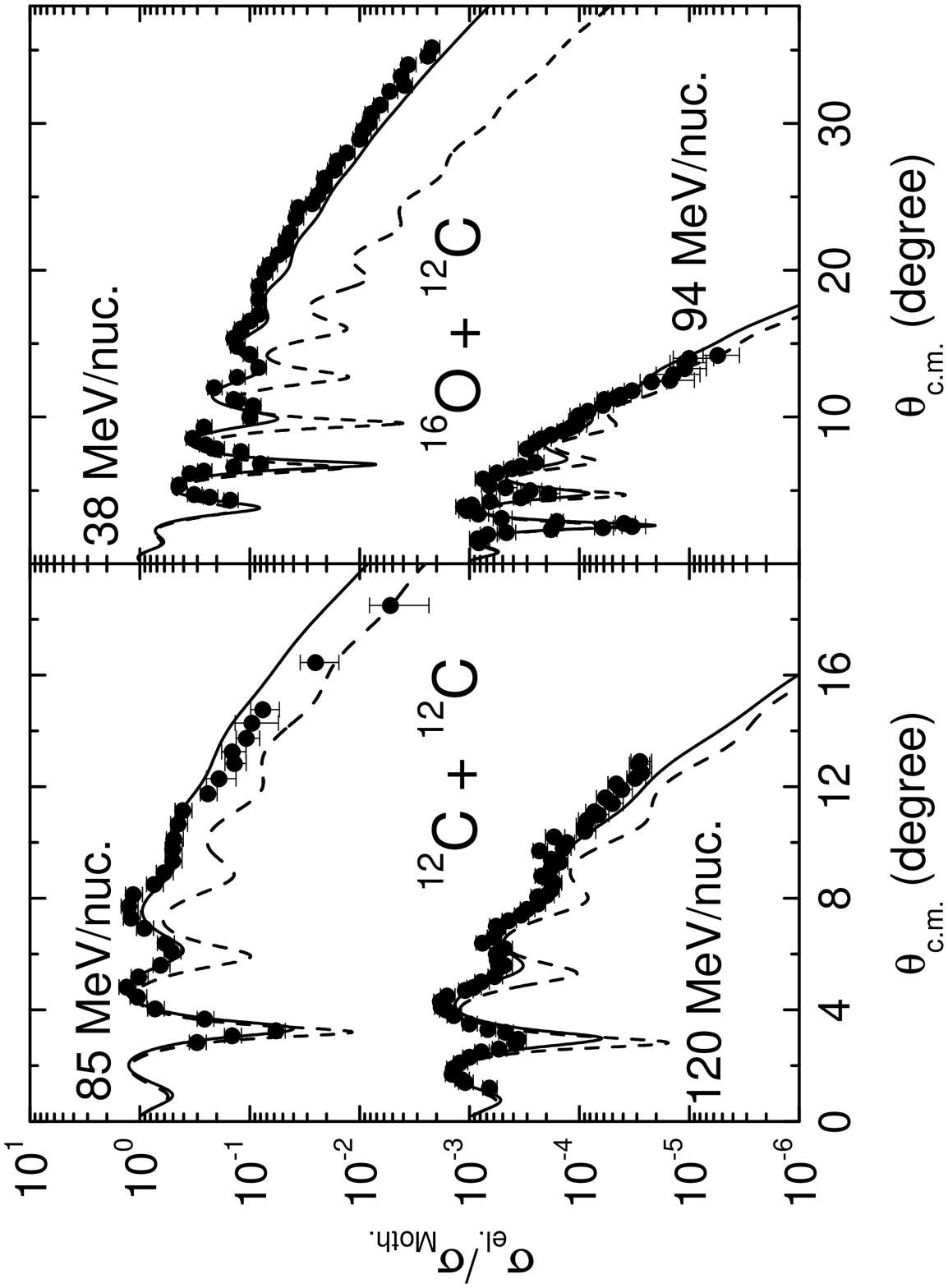}
\vspace{7.0cm}
\caption{The same as Fig. 1, for the $^{12}$C,$^{16}$O + $^{12}$C systems.}
\end{figure}

\begin{figure}
\vspace{0.0cm}
\hspace{-10.0cm}
\includegraphics{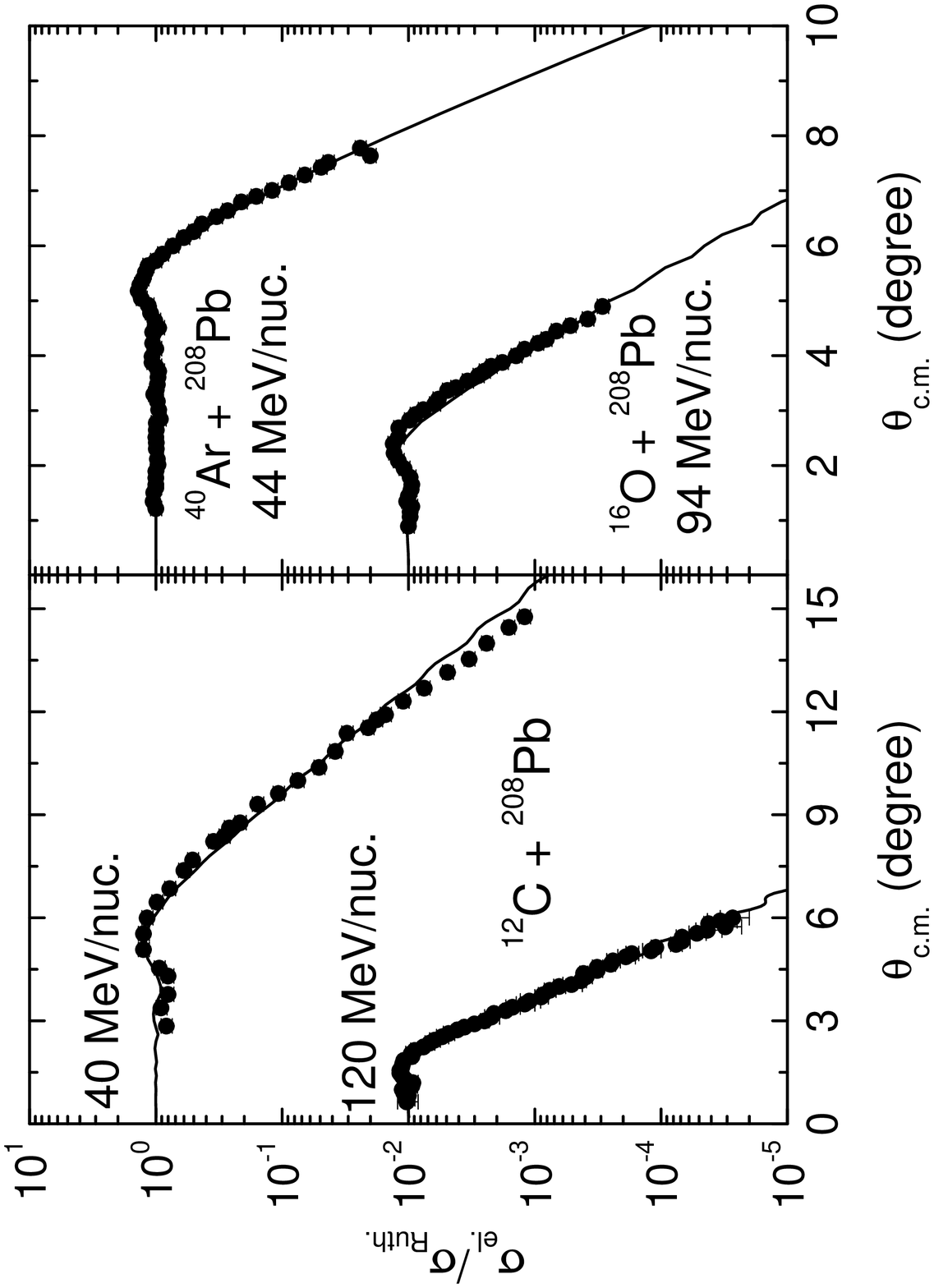}
\vspace{7.0cm}
\caption{The same as Fig. 1, for the $^{12}$C,$^{16}$O,$^{40}$Ar + $^{208}$Pb 
systems.}
\end{figure}

\begin{figure}
\vspace{0.0cm}
\hspace{-10.0cm}
\includegraphics{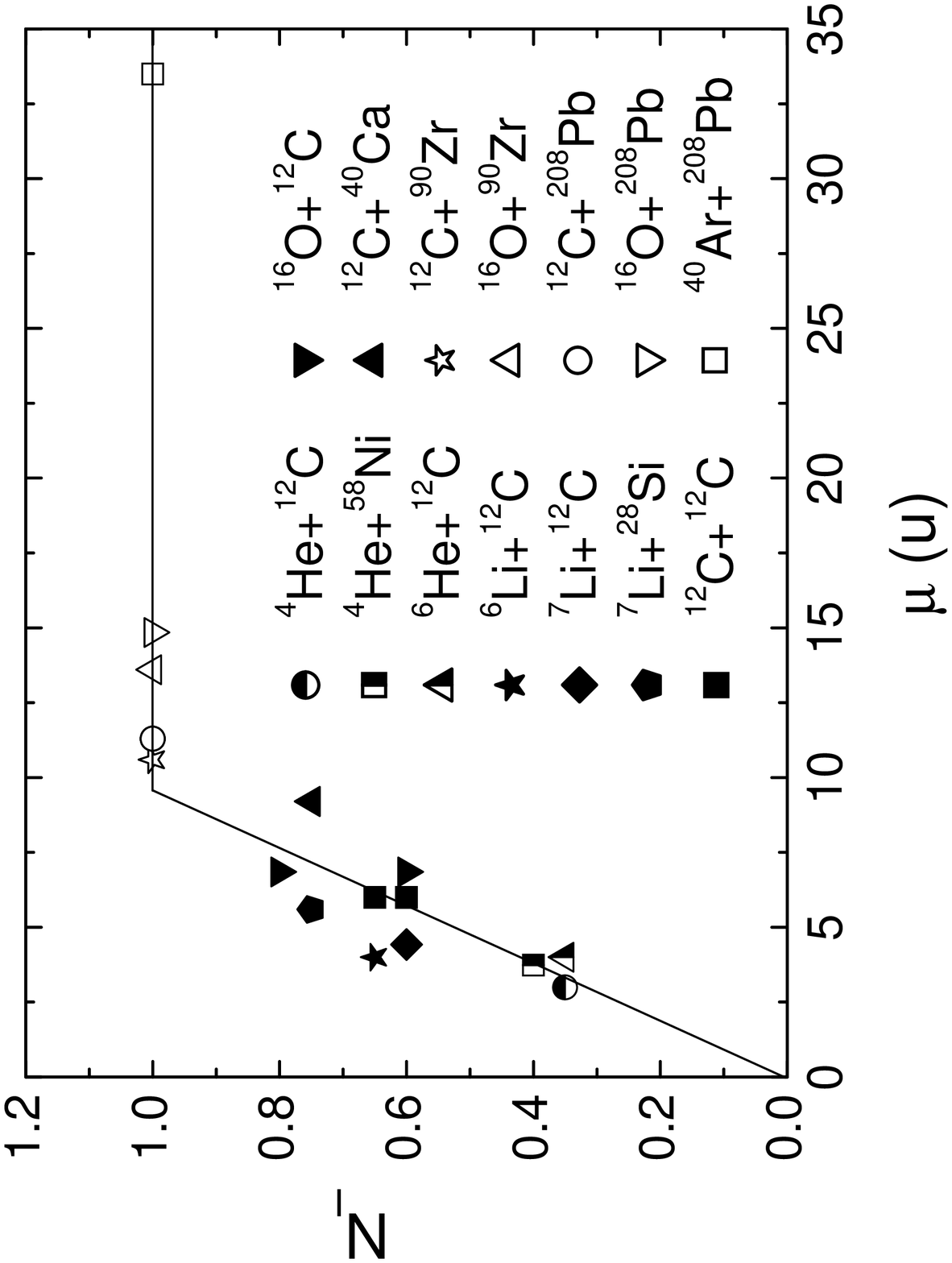}
\vspace{8.0cm}
\caption{The factor of normalization of the imaginary part of the potential as 
a function of the reduced mass of the system. The solid lines are just guides.}
\end{figure}

\begin{figure}
\vspace{10.0cm}
\hspace{-9.0cm}
\includegraphics{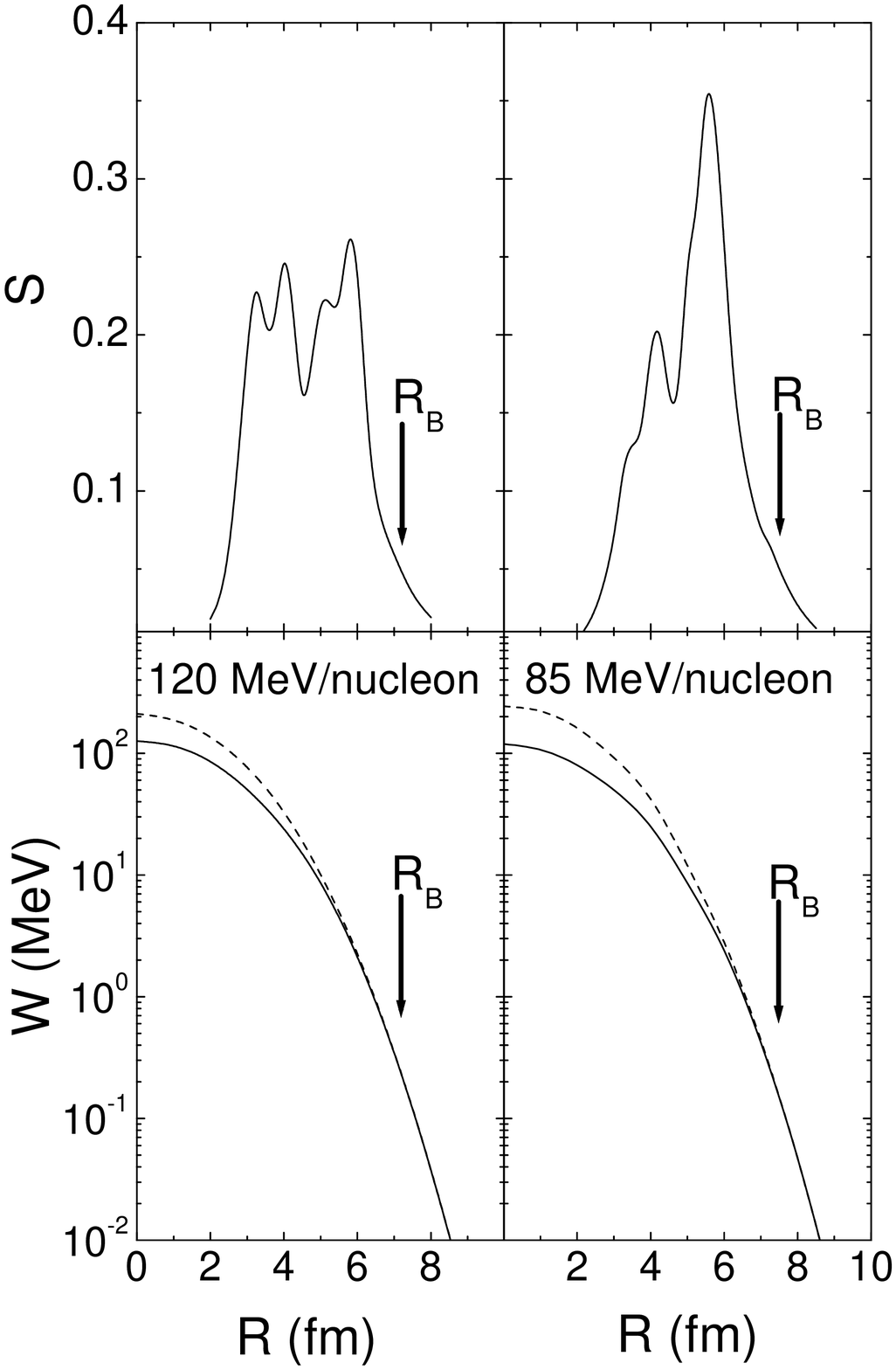}
\vspace{0.0cm}
\caption{Bottom - The imaginary part of the optical potential with (solid lines)
or without (dashed lines) including the effect of Pauli blocking, for the
$^{12}$C + $^{12}$C system in two different bombarding energies. The arrows 
indicate the positions of the s-wave barrier radii. Top - The results of the 
notch test corresponding to these two elastic scattering angular distributions.
The quantity $S$ represents the relative difference between the chi-squares
obtained with and without a Gaussian perturbation, centered in $R$, in the 
imaginary part of the potential.}
\end{figure}

\begin{figure}
\vspace{0.0cm}
\hspace{-10.0cm}
\includegraphics{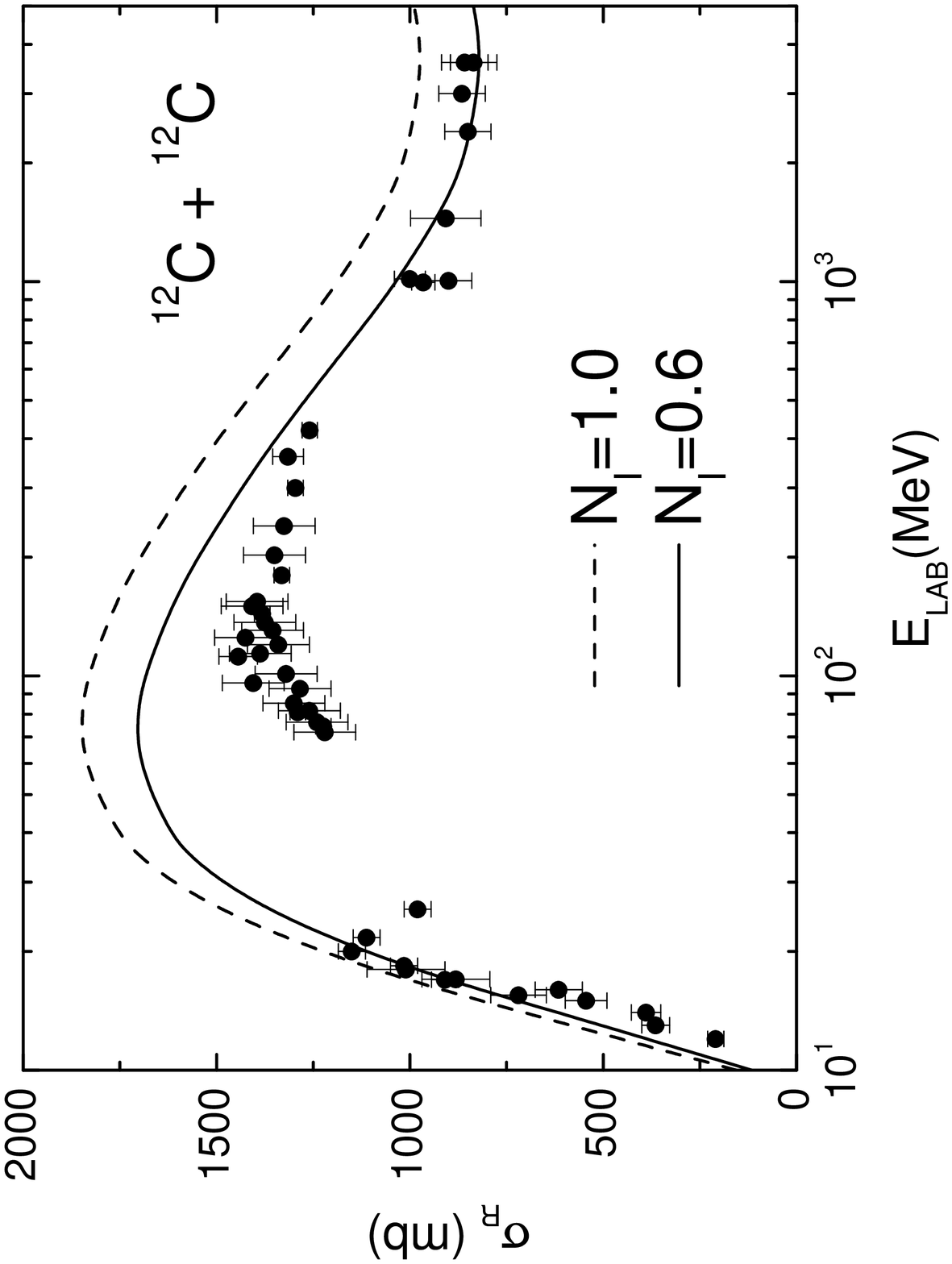}
\vspace{7.0cm}
\caption{The reaction cross section as a function of the energy for the 
$^{12}$C + $^{12}$C system. The lines represent the results obtained from 
optical model calculations, with different factors of normalization for the 
imaginary part of the potential.}
\end{figure}

\begin{figure}
\vspace{10.0cm}
\hspace{-9.0cm}
\includegraphics{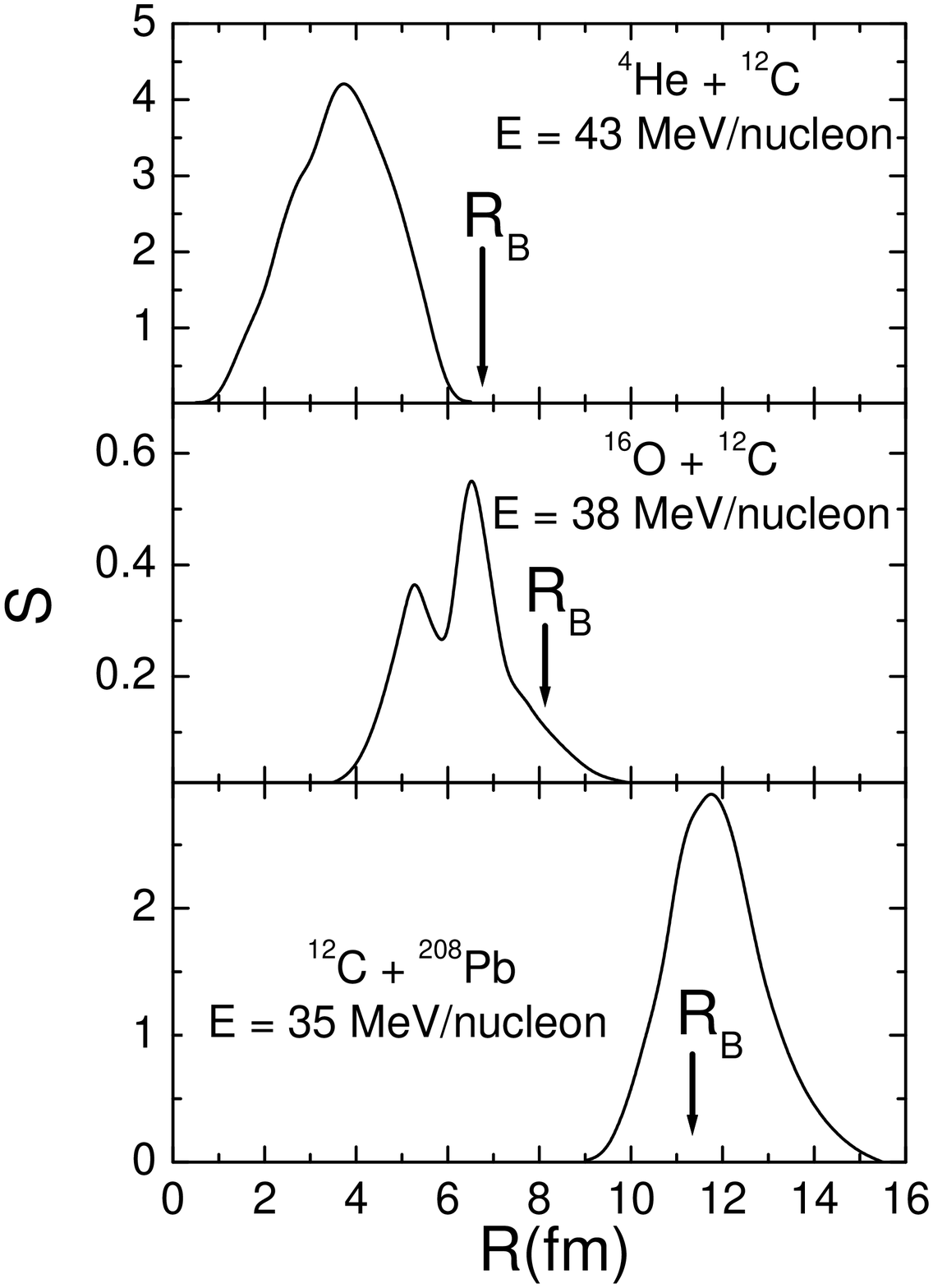}
\vspace{-1.0cm}
\caption{The regions of sensitivity for three different systems as determined by
the notch test. The arrows indicate the positions of the s-wave barrier radii.}
\end{figure}

\newpage

\begin{table}
\caption{Systems that have been analyzed in the present work, their 
corresponding reduced masses, and the energies (in MeV/nucleon units) of the 
respective elastic scattering angular distributions. The table also presents 
the best fit values for the factor of normalization $N_I$ of the imaginary part 
of the optical potential, the volume integrals ($J_R$ in MeV fm$^3$ units) for 
the real part of the optical potential, and the reaction cross sections
($\sigma_R$ in mb units) obtained in the present work (PW). For purpose of 
comparison, the ranges for $J_R$ and $\sigma_R$ obtained in earlier works (EW -
see refs. in the text) are also included in the table.} 
\begin{ruledtabular}
\begin{tabular}{ccccccccc}
Proj.&Target&E&$\mu$(u)&$N_I$&$J_{R}$ PW&$J_{R}$ EW&$\sigma_{R}$ PW&$\sigma_{R}$
EW \\
\hline
$^4$He&$^{12}$C&  43 & 3.00 & 0.35&282&244-277&659&718\\
$^4$He&$^{58}$Ni&  43 & 3.74 & 0.40&274&251-287&1437&\\
$^6$He&$^{12}$C&  38 & 4.00 & 0.35&299&&923&1092-1179\\
$^6$Li&$^{12}$C&  53 & 4.00 & 0.65&267&239-296&964&904-1184\\
$^7$Li&$^{12}$C&  50 & 4.42 & 0.60&271&260-288&1003&969-1021\\
$^7$Li&$^{28}$Si&  50 & 5.60 & 0.75&267&227-257&1491&1462-1666\\
$^{12}$C&$^{12}$C&  85 & 6.00 & 0.60&206&185&1061&1000\\
$^{12}$C&$^{12}$C& 120 & 6.00 & 0.65&156&140&979&907\\
$^{16}$O&$^{12}$C&  38 & 6.85 & 0.60&296&225-233&1463&1374\\
$^{16}$O&$^{12}$C&  94 & 6.85 & 0.80&191&156&1227&1136\\
$^{12}$C&$^{40}$Ca&  25 & 9.20 & 0.75&323&&2209&2030\\
$^{12}$C&$^{90}$Zr&  25 & 10.60 & 1.00&320&&2993&2415\\
$^{12}$C&$^{90}$Zr&  35 & 10.60 & 1.00&297&&2939&2840\\
$^{12}$C&$^{208}$Pb&  25 & 11.30 & 1.00&320&&3820&3300\\
$^{12}$C&$^{208}$Pb&  35 & 11.30 & 1.00&298&&3894&3561\\
$^{12}$C&$^{208}$Pb&  40 & 11.30 & 1.00&287&&3898&\\
$^{12}$C&$^{208}$Pb& 120 & 11.30 & 1.00&158&&3629&3136\\
$^{16}$O&$^{90}$Zr&  94 & 13.58 & 1.00&190&&2752&2749\\
$^{16}$O&$^{208}$Pb&  94 & 14.85 & 1.00&192&&3879&3485\\
$^{40}$Ar&$^{208}$Pb&  44 & 33.50 & 1.00&280&&5017&\\
\end{tabular}
\end{ruledtabular}
\end{table}

\begin{table}
\caption{Values obtained for the diffuseness of the matter densities. The
diffuseness indicated for the $^4$He nucleus was obtained from the corresponding
charge density. The root-mean-square radii of the matter ($RMS_M$) and charge 
($RMS_C$) distributions are included in the table.} 
\begin{ruledtabular}
\begin{tabular}{cccc}
Nucleus&$a$ (fm)&$RMS_M$ (fm)&$RMS_C$ (fm)\\
\hline
$^4$He&$\approx$ 0.3&1.68&1.68\\
$^6$He&0.56&2.40&\\
$^6$Li&0.56&2.40&2.55\\
$^7$Li&0.54&2.38&2.39\\
$^{12}$C&0.58&2.73&2.47\\
$^{16}$O&0.59&2.90&2.74\\
$^{28}$Si&0.59&3.27&3.10\\
$^{40}$Ca&0.56&3.50&3.48\\
$^{40}$Ar&0.56&3.50&3.41\\
$^{58}$Ni&0.59&3.94&3.77\\
$^{90}$Zr&0.56&4.42&4.26\\
$^{208}$Pb&0.54&5.72&5.50\\
\end{tabular}
\end{ruledtabular}
\end{table}


\begin{thebibliography}{99}
\bibitem{nova4} M. E. Brandan and G. R. Satchler, Phys. Rep. {\bf 285} (1997) 
143.
\bibitem{4} L. C. Chamon, D. Pereira, M. S. Hussein, M. A. Candido Ribeiro and 
D. Galetti, Phys. Rev. Lett. {\bf 79} (1997) 5218.
\bibitem{10} L. C. Chamon, B. V. Carlson, L. R. Gasques, D. Pereira, C. De Conti,
M. A. G. Alvarez, M. S. Hussein, M. A. Candido Ribeiro, E. S. Rossi Jr. and C. 
P. Silva, Phys. Rev. C {\bf 66} (2002) 014610.
\bibitem{15} M. S. Hussein, R. A. Rego and C. A. Bertulani, Phys. Rep. {\bf 201}
(1991) 279.
\bibitem{1} F. Perey and B. Buck, Nucl. Phys. {\bf 32} (1962) 253.
\bibitem{2} D. F. Jackson and R. C. Johnson, Phys. Lett. {\bf B49} (1974) 249.
\bibitem{3} M. A. Candido Ribeiro, L. C. Chamon, D. Pereira, M. S. Hussein and 
D. Galetti, Phys.Rev.Lett. {\bf 78} (1997) 3270.
\bibitem{5} L. C. Chamon, D. Pereira and M. S. Hussein, Phys. Rev. C {\bf 58} 
(1998) 576. 
\bibitem{6} M. A. G. Alvarez, L. C. Chamon, D. Pereira, E. S. Rossi Jr., C. P.
Silva, L. R. Gasques, H. Dias and M. O. Roos, Nucl. Phys. {\bf A656} (1999) 187.
\bibitem{7} L. R. Gasques, L. C. Chamon, C. P. Silva, D. Pereira, M. A. G.
Alvarez, E. S. Rossi Jr., V. P. Likhachev, B. V. Carlson and C. De Conti, Phys.
Rev. {\bf C65} (2002) 044314.
\bibitem{8} M. A. G. Alvarez et al., Phys. Rev. {\bf C65} (2002) 014602.
\bibitem{9} E. S. Rossi Jr., D. Pereira, L. C. Chamon, M. A. G. Alvarez, L. R. 
Gasques, J. Lubian, B. V. Carlson and C. De Conti, Nucl.\ Phys.\ {\bf A707} 
(2002) 325.
\bibitem{11} L. R. Gasques, L. C. Chamon, D. Pereira, M. A. G. Alvarez, E. S.
Rossi Jr., C. P. Silva, G. P. A. Nobre and B. V. Carlson, Phys. Rev. {\bf C67}
(2003) 067603.
\bibitem{12} L. R. Gasques et al, Phys. Rev. {\bf C67} (2003) 024602.
\bibitem{13} M. A. G. Alvarez, L. C. Chamon, M. S. Hussein, D. Pereira, E. S. 
Rossi Jr. and C. P. Silva, Nucl. Phys. {\bf A723} (2003) 93.
\bibitem{14} L. R. Gasques, L. C. Chamon, D. Pereira, M. A. G. Alvarez, E. S.
Rossi Jr., C. P. Silva and B. V. Carlson, Phys. Rev {C69} (2004) 034603.
\bibitem{16} R. J. Glauber, W. E. Brittin, L. G. Dunhan (Eds.), Lectures in 
Theoretical Physics,{\bf Vol. 1}, Wiley-Interscience, New York, 1959, p. 315;
R. J. Glauber, in: G. Alexander (Ed.), High Energy Physics and 
Nuclear Structure, Wiley, New York, 1967, p. 311; R. J. Glauber High Energy 
Physics and Nuclear Structure, Plenum, New York, 1970, p.207.
\bibitem{nova2} Wilmot N. Hess, Rev. Mod. Phys. {\bf 30} (1958) 368.
\bibitem{20} J. Y. Hostachy, M. Buerned, J. Chauvin, D. Lebrun, Ph. Martin,
 J. C. Lugol, L. Papineau, P. Roussel, N. Alamanos, J. Arvieux and
C. Cerruti, Nucl.\ Phys.\ {\bf A490} (1988) 441.
\bibitem{21} M. E. Brandan, H. Chehime and K. W. Mc Voy, Phys.\ Rev.\ C {\bf 55}, 1353 (1997).
\bibitem{22} V. Lapoux, et al.,  Phys. \ Rev. \ C {\bf 66} (2002) 034608.
\bibitem{23} A. Nadasen et al, Phys.\ Rev.\ C {\bf 52} (1995) 1894.
\bibitem{24} C. C. Sahm, T. Murakami, J. G. Cramer, A. J. Lazzarini, D. D. 
Leach, D. R. Tieger, R. A. Loveman, W. G. Lynch, M. B. Tsang and J. Van der 
Plicht, Phys.\ Rev.\ C {\bf 34} (1986) 2165.
\bibitem{25} Y. T. Oganessian, Y. E. Penionzhkevich, V. I. Man'Ko and 
V. N. Polyansky, Nucl.\ Phys.\ {\bf A303} (1978) 259. 
\bibitem{26} P. Roussel-Chomaz, N. Alamanos, F. Auger, J. Barrette, B. Berthier,
B. Fernandez, L. Papineau, Nucl.\ Phys.\ {\bf A477} (1988) 345. 
\bibitem{27} C. Olmer, M. Mermaz, M. Buenerd, C. K. Gelbke, D. L. Hendrie, 
J. Mahoney, D. K. Scott, M. H. Macfarlane, S. C. Pieper, Phys. Rev.
{\bf C18} (1978) 205.
\bibitem{28} J. Albinski, A. Budzanowski, H. Dabrowski, Z. Rogalska, S. Wiktor, H. Rebel, D. K. Srivastava, C. Alderliesten, J. Bojowald, W.
Oelert, C. Mayer-Boricke and P. Turek, Nucl. Phys. {\bf A445} (1995) 477.
\bibitem{19} H. De Vries, C. W. De Jager and C. De Vries, Atomic Data and Nucl.
Data Tables {\bf 36} (1987) 495.
\bibitem{29} C. Perrin, S. Kox, N. Longequeue, J. B. Viano, M. Buenerd, R. 
Cherkaoui, A. J. Cole, A. Gamp, J. Menet, R. Ost, R. Bertholet, C.
Guet and J. Pinston, Phys. Rev. Lett. {\bf 49} (1982) 1905. 
\bibitem{30} A. J. Cole, W. D. M. Rae, M. E. Brandan, A. Dacal, B. G. Harvey, 
R. Legrain, M. J. Murphy and R. G. Stokstad, Phys. Rev. Lett. {\bf 47} 
(1981) 1705.
\bibitem{31} S. Kox, A. Gamp, R. Cherkaoni, A. J. Cole, N. Longequeue, J. Menet, 
C. Perin and J. B. Viano, Nucl. Phys. {\bf A420} (1984) 162.
\bibitem{32} S. Kox et al, Phys. Rev. C {\bf 35} (1987) 1678.
\bibitem{33} H. G. Bohlen, M. R. Clover, G. Ingold, H. Lettan, W. von Oertzen, 
Z. Phys. {\bf A308} (1982) 121.
\bibitem{34} M. Buerned, A. Lounis, J. Chauvin, D. Lebrun, P. Martin, 
G. Duhamel, J. C. Gondrand, P. de Saintgnon, Nucl. Phys. {\bf A424} (1994) 313.
\bibitem{37} A. A. Ogloblin et al, Phys. Rev. {\bf C62} (2000) 44601 .
\bibitem{38} M. H. Cha, and Y. J. Kim, Phys. Rev. {\bf C51} (1995) 212.
\bibitem{39} D. T. Khoa, G. R. Satchler, and W von Oertzen, Phys. Rev. 
{\bf C51} (1995) 2069.
\bibitem{40} D. T. Khoa, W von Oertzen, and H. G. Bohlen, Phys. Rev. {\bf C49},
(1994) 1652.
\bibitem{42} D. T. Khoa and G. R. Satchler, Nucl. Phys. {\bf A668} (2000) 3.
\bibitem{43} M. El-Azab Farid and M. A. Hassanain, Nucl. Phys. {\bf A678} (2000)
39. 
\bibitem{44}  D. T. Khoa, G. R. Satchler, and W von Oertzen, Phys. Rev. 
{\bf C56} (1997) 954.
\end{thebibliography}
\end{document}